\documentclass[prl,twocolumn,showpacs,floats,amsmath,amssymb]{revtex4}
\usepackage[dvips]{graphicx}
\begin{document}
\title{Pumping $ac$ Josephson current in the Single Molecular Magnets by spin nutation}
\author{B. Abdollahipour$^{1*}$, J. Abouie$^{2,3}$\footnote{These authors have contributed equally to this work.},
and A. A. Rostami$^4$}
\affiliation{$^1$Department of Physics, University of Tabriz, Tabriz 51666-16471, Iran\\
$^2$ Department of Physics, Institute for Advanced Studies in Basic Sciences (IASBS), Zanjan 45137-66731, Iran\\
$^3$ School of physics, Institute for Research in Fundamental Sciences (IPM),
Tehran 19395-5531, Iran\\
$^4$ Department of Physics, Shahrood University of Technology, Shahrood 36199-95161,
Iran}

\begin{abstract}
We demonstrate that an {\it ac} Josephson current is pumped through
the Single Molecular Magnets (SMM) by the spin nutation. The spin
nutation is generated by applying a time dependent magnetic field to
the SMM. We obtain the flowing charge current through the junction
by working in the tunneling limit and employing Green's function
technique. At the resonance conditions some discontinuities and
divergencies are appeared in the normal and Josephson currents,
respectively. Such discontinuities and divergencies reveal
themselves when the absorbed/emitted energy, owing to the
interaction of the quasiparticles with the spin dynamics are in the
range of the superconducting gap.
\end{abstract}
\maketitle

%%%%%%%%%%%%%%%%%%%%%%%%%%%%%%%%%%%%%%%%%%%%%%%%%%%%%%%%%%%%%%%%%%%%%
%\section{Introduction}
%%%%%%%%%%%%%%%%%%%%%%%%%%%%%%%%%%%%%%%%%%%%%%%%%%%%%%%%%%%%%%%%%%%%%

Quantum pumping is a coherent transport mechanism to produce a
charge current in the absence of an external bias voltage by an
appropriate periodical variation of the system parameters
\cite{Thouless1983,Buttiker1994,Brouwer1998,Zhou1999}. It has been
introduced as a potential way to generate a dissipationless charge
current in the nanoelectronic devices \cite{Altshuler1999}. The
adiabatic quantum pumping is also a method for generating a
dynamically controlled flow of spin-entangled electrons, which is
promising because of the vast expertise already available in
solid-state electronics \cite{Das2006}. In recent years, electron
pumps consisting of different systems such as small semiconductor
quantum dots \cite{Pothier92, Matinis94,Switkes1999}, carbon
nanotube quantum dot \cite{Buitelaar2008}, one-dimensional
interacting L\"{u}ttinger liquid quantum wire \cite{Sharma2001},
Josephson junctions with half-metallic ferromagnets
\cite{Takahashi2007} and diffusive ferromagnets \cite{Houzet2008},
and InAs nanowire embedded in a superconducting quantum interference
device (SQUID) \cite{Giazotto2011} have received considerable
theoretical and experimental attentions. Several different
mechanisms have been proposed to pump charge through such systems,
ranging from a low-frequency modulation of gate voltages in
combination with the Coulomb blockade to photon-assisted transport.

Magnetic Josephson junctions consisting of Single Molecular Magnets
(SMM) and magnetic nanoparticles have recently attracted intense
attentions, owing to their applications in molecular spintronics
devices \cite{Lapo08} and classical \cite{Leue01} and quantum
information processing \cite{Affr09}. The small sizes of these
junctions are an advantage for their application. Compounds of single
molecular magnet class \cite{Chri00, Gatt07} are particularly
attractive for application in high-density information storage and
quantum computing, due to their long magnetization relaxation time
at low temperatures \cite{Carr07, Arda07}. The rich physics behind
the magnetic behavior produces interesting effects such as negative
differential conductance and complete current suppression
\cite{Jo2006,Heersche2006}, which could be used in nanoelectronics.
%
%%%%%%%%%%%%%%%%%%%%%%%%%%%%%%%%%%%%%%%%%%%%%%%%%%%%%%%%%%%%%%%%%%%%
%%%%%%%%%%%%%%%%%%%%%      Fig 1     %%%%%%%%%%%%%%%%%%%%%%%%%%%%%%%
\begin{figure}[h]
\centerline{\includegraphics[width=7cm]{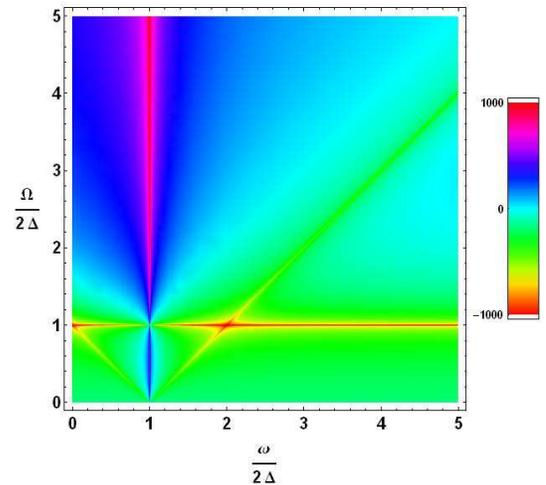}} \caption{(Color
online) Color map of the amplitude of the $ac$ Josephson current
$\bar{\cal {J}}$ as a function of $\frac{\omega}{2\Delta}$ and
$\frac{\Omega}{2\Delta}$. At $\frac{\Omega}{2\Delta}\sim 1$ the
amplitude goes down to very small values (infinity) for
$\frac{\omega}{2\Delta}\neq 1$. At $\frac{\omega}{2\Delta}\sim 1$
the amplitude diverges for $\frac{\Omega}{2\Delta}\neq 1$.}
\label{amplitude-josephson}
\end{figure}
%%%%%%%%%%%%%%%%%%%%%%%%%%%%%%%%%%%%%%%%%%%%%%%%%%%%%%%%%%%%%%%%%%%%
%

In this letter, we show that the spin nutation of a SMM could pump
the charge current through the Josephson junction consisting of the
SMM. The spin nutation can be generated by applying an external time
dependent magnetic field to the SMM which is a combination of a
static and a rotating transverse {\it rf} fields. We consider a SMM
connected to the two spin-singlet superconducting (SSC) leads via
tunnel barriers. We investigate the coupling of the spin nutation
and Josephson current through the junction. Interplay of the
Josephson current and a precessing spin between various types of
superconducting leads connected via tunnel barriers has been
considered by Zhu and Balatsky \cite{Zhu2003}. The Josephson current
through the junction consisting of two SSC leads is not modulated by
the spin precession. Whereas, when both superconductors have equal
spin triplet pairing state, spin precession causes to modulate the
Josephson current with twice the Larmor frequency. In addition, it
has been shown that a circularly polarized {\it ac} spin current
with the Larmor frequency is generated in the SSC leads in response
to the spin precession \cite{Teber2010}.

Working in the tunneling limit and employing normal and anomalous
Green's functions we obtain the flowing current through the
junction. We have found that the normal current, the current
associated to the single particle tunneling, and Josephson current
are modulated by spin nutation. In response to the time-dependent
boundary conditions induced by spin nutation, $ac$ normal and
Josephson currents are pumped in the junction. Some discontinuities
and divergencies are appeared in the amplitudes of the $ac$ normal
and $ac$ Josephson currents, respectively (Fig.
\ref{amplitude-josephson}). Such discontinuities and divergencies
reveal themselves when the absorbed or emitted energy of charge
carriers, owing to the interaction with the spin dynamics are in the
range of the superconducting gap. At these situations the resonance
conditions are fulfilled by the system.

%%%%%%%%%%%%%%%%%%%%%%%%%%%%%%%%%%%%%%%%%%%%%%%%%%%%%%%%%%%%%%%%%%%%
%%%%%%%%%%%%%%%%%%%%%%%%   Model   %%%%%%%%%%%%%%%%%%%%%%%%%%%%%%%%%
%%%%%%%%%%%%%%%%%%%%%%%%%%%%%%%%%%%%%%%%%%%%%%%%%%%%%%%%%%%%%%%%%%%%

\begin{figure}[h]
\centerline{\includegraphics[width=4cm]{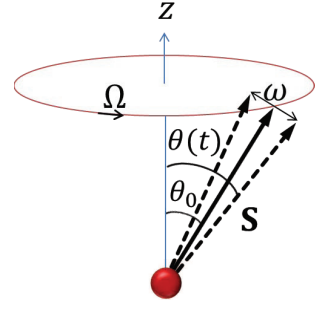}}
\caption{Schematic representation of SMM spin nutation.}
\label{spin-nutation}
\end{figure}

{\it Model} - Let us consider the ${\rm SSC|SMM|SSC}$ Josephson
junction. By considering the SMM as a classical spin vector
$(\mathbf S)$, and applying the time dependent magnetic field
$\mathbf{h}(t)=\left(-h_0\sin\omega t~\sin\Omega t, h_0\sin\omega
t~\cos\Omega t, h_z\right)$ to the SMM, in the absence of the spin
relaxation processes the dynamics of the spin is given by:
\begin{equation}\label{spin-smm}
\mathbf{S}(t)=S\left(\sin\theta(t)\cos\varphi(t),\sin\theta(t)\sin\varphi
(t),\cos\theta(t)\right).
\end{equation}
Where
\begin{equation}\label{theta&phi}
\varphi(t)=\Omega t,\;\;\theta(t)=\theta_0-\vartheta\cos\omega t,
\end{equation}
$\Omega=\gamma h_z$ ($\gamma$ denotes gyromagnetic ratio) is the
precession frequency around the $z$ axis and $\theta(t)$ is the time
dependent tilt angle (angle between the spin and $z$ axis). The tilt
angle oscillates about $\theta_0$ with frequency $\omega$ and
amplitude $\vartheta=\gamma h_0/\omega$. (See
Fig.(\ref{spin-nutation})) Indeed, this nutational motion of the
spin is served as the pumping parameters in the system.
%
%%%%%%%%%%%%%%%%%%%%%%%%%%%%%%%%%%%%%%%%%%%%%%%%%%%%%%%%%%%%%%%%%%%%
%%%%%%%%%%%%%%%%%%%%%      Fig 2     %%%%%%%%%%%%%%%%%%%%%%%%%%%%%%%
%%%%%%%%%%%%%%%%%%%%%%%%%%%%%%%%%%%%%%%%%%%%%%%%%%%%%%%%%%%%%%%%%%%%
%
Let us consider the following Hamiltonian for the Josephson
junction:
\begin{equation}\label{total-hamiltonian}
H(t)=H_L+H_R+H_T(t)\ .
\end{equation}
The first two terms describe the energy of the left ($L$) and right
($R$) spin singlet superconducting leads and are given by
\begin{equation}\label{hamiltonian-leads}
H_{\alpha}=\sum_{k,\sigma=\uparrow,\downarrow}\varepsilon_k
c_{\alpha,k,\sigma}^{\dagger}c_{\alpha,k,\sigma}+\sum_k\left(\Delta_{\alpha}
c_{\alpha,k,\uparrow}^{\dagger}c_{\alpha,-k,\downarrow}+h.c.\right)
\end{equation}
where $c_{\alpha,k,\sigma}^{\dagger}(c_{\alpha,k,\sigma})$ is the
creation (annihilation) operator of an electron on the lead
$\alpha=L, R$ with momentum $k$ and spin $\sigma$. $\varepsilon_k$
is the energy of single conduction electron and
$\Delta_{\alpha}=\Delta e^{i\chi_{\alpha}}$ is the pair potential in
which $\chi_{\alpha}$ is the superconducting phase of the lead
$\alpha$. The two leads are weakly coupled via the tunneling
Hamiltonian, $H_T(t)$ which is given by
\begin{equation}\label{tunneling-hamiltonian}
H_T(t)=\sum_{k,k',\sigma\sigma'}\left(c_{R,k,\sigma}^{\dagger}
T_{\sigma\sigma'}(t)c_{L,k',\sigma'}+h.c.\right)\,
\end{equation}
$T_{\sigma\sigma'}$ is a component of the following time dependent
tunneling matrix
\begin{equation}\label{tunneling-matrix}
\hat{T}(t)=T_0\hat{\mathbf{1}}+T_S\hat{\mathbf{S}}(t)\cdot\hat{\mbox{\boldmath
$\sigma$}}\ ,
\end{equation}
where $\hat{\mathbf{1}}$ is the $2\times 2$ unit matrix,
$\hat{\mbox{\boldmath $\sigma$}}=(\sigma_x,\sigma_y,\sigma_z)$ is
the Pauli matrices and
$\hat{\mathbf{S}}(t)=\frac{\mathbf{S}}{|\mathbf{S}|}$ is the unit
vector along the SMM spin direction. $T_0$ is the direct spin
independent transmission amplitude and $T_S$, which is originated
from the exchange interaction between the spin of conduction
electrons and the localized spin of SMM, indicate the spin dependent
transmission amplitude.

In the tunneling limit and zero bias voltage, we could formally
separate the current operator at lead $\alpha$ in two parts
\cite{Mahan1990}
\begin{equation}\label{current-operator}
I_{\alpha}(t)=I_{\alpha}^s(t)+I_{\alpha}^J(t)
\end{equation}
where
$I_{\alpha}^s(t)=-e\int_{-\infty}^{t}dt'\left(\left\langle\left[A_{\alpha}(t),
A_{\alpha}^{\dagger}(t')\right]\right\rangle+h.c.\right)$ and
$I_{\alpha}^J(t)=-e\int_{-\infty}^{t}dt'\left(\left\langle\left[A_{\alpha}(t),
A_{\alpha}(t')\right]\right\rangle+h.c.\right)$ are the normal
current and Josephson current carried by single particles and Cooper
pairs, respectively. The operator $A_{\alpha}(t)$, is given by
\begin{equation}\label{a-operator}
A_{\alpha}(t)=\sum_{k,k',\sigma\sigma'}c_{\alpha',k,\sigma}^{\dagger}(t)
T_{\sigma\sigma'}(t)c_{\alpha,k',\sigma'}(t)\ ,
\end{equation}
where $\alpha=L,R$ and $\alpha'=R,L$.

%%%%%%%%%%%%%%%%%%%%%%%%%%%%%%%%%%%%%%%%%%%%%%%%%%%%%%%%%%%%%%%%%%%%
%%%%%%%%%%%%%%%%%%%%%%%%   Normal Current     %%%%%%%%%%%%%%%%%%%%%%

{\it Normal current} - In the Following we will show that the spin
nutation causes to transfer electrons through the junction and a
time-dependent normal current emerges. Let us define the following
retarded potential
\begin{equation}\label{U-retarded}
U^{\sigma\sigma'}_{\rho\rho'}(t-t')=-i\Theta(t-t')\left\langle\left[a_{k,k'}^{\sigma\sigma'}(t),
a_{p,p'}^{\dagger\rho\rho'}(t')\right]\right\rangle\ ,
\end{equation}
where, $a_{k,k'}^{\sigma\sigma'}(t)=c_{\alpha,k,\sigma}^{\dagger}(t)
c_{\alpha,k',\sigma'}(t)$ and
$a_{p,p'}^{\rho\rho'}(t')=c_{\alpha,p,\rho}^{\dagger}(t')
c_{\alpha,p',\rho'}(t')$. Since there is no spin dependent
interaction inside the superconducting leads, the retarded potential
(\ref{U-retarded}) is independent on spin indices and
$U^{\sigma\sigma'}_{\rho\rho'}(t-t')=\delta_{\sigma\rho}\delta_{\sigma'\rho'}U_{ret}(t-t')$.
The Fourier transformation of the Matsubara potential, with
imaginary time, is defined as \cite{Mahan1990}
\begin{equation}\label{U-Matsobara}
\mathcal{U}(i\omega_n)=\frac{1}{\beta}\sum_{iq}g_R(k,iq-i\omega_n)g_L(k',iq)\ ,
\end{equation}
where $g_R$ and $g_L$ are the normal Matsubara Green's functions and
$\omega_n$'s indicate the Matsubara frequencies. The retarded
potential is obtained from the Matsubara potential by analytical
continuation $i\omega_n\rightarrow\alpha+i\eta$ where
$\eta\rightarrow 0^+$. Employing Lehman representation we can
calculate the retarded potential and obtain the normal current
\begin{equation}
i^s(t)=\Delta T_{\bot} \left[2T_{\bot}{\cal
S}\left(\frac{\Omega}{2\Delta}\right) + T_{||}\bar{{\cal
S}}\left(\frac{\Omega}{2\Delta},
\frac{\omega}{2\Delta}\right)\vartheta\cos\omega t\right],
\label{normal-current}
\end{equation}
where $i^s=I^s/2\pi eN_LN_R$ and $N_L(N_R)$ is the density of states
at the Fermi energy in the left (right) lead. To obtain this
equation we have considered $\vartheta/\theta_0\ll 1$. This
approximation is fulfilled by the practical conditions of the system
and dose not make any restriction on it. The parameters
$T_{||}=T_S\cos\theta_0$ and $T_{\bot}=T_S\sin\theta_0$ are spin
conserving and spin-flip transmission amplitudes, respectively. As
it is clearly seen, the normal current strongly depends on the
spin-flip transmission amplitude. If $T_{\bot}$ is zero  the normal
current vanishes completely. This situation corresponds to the
oscillation of the SMM around $\theta_0=n\pi~(n=0,1,\dots)$. If the
spin-flip transmission amplitude is nonzero ($T_{\bot}\neq 0$),
depending on the strength of precession frequency $\Omega$ and tilt
angle oscillation frequency $\omega$, the single particles could be
transferred from one lead to another by absorbing and emitting a
quantum of oscillation. In this case the normal current depends on
the parameters $\bar{{\cal S}}\left(\frac{\Omega}{2\Delta},
\frac{\omega}{2\Delta}\right)$ and ${\cal
S}\left(\frac{\Omega}{2\Delta}\right)$, which are given by
\begin{eqnarray}\label{M-definition}
&\nonumber\bar{{\cal S}}\left(\frac{\Omega}{2\Delta}, \frac{\omega}{2\Delta}\right)=
\nonumber 2{\cal S}\left(\frac{\omega}{2\Delta}\right)-2{\cal S}\left(\frac{\Omega}{2\Delta}\right)
-{\cal S}\left(\frac{\Omega+\omega}{2\Delta}\right)\\
&-{\cal S}\left(\frac{\Omega-\omega}{2\Delta}\right)-{\cal S}\left(\frac{\omega-\Omega}{2\Delta}\right),\\
&\nonumber {\cal S}(x)=\Theta(x-1)
\left\{\frac{x^2}{x+1}K(\gamma)
-(x+1)[K(\gamma)-E(\gamma)]\right\},
\end{eqnarray}
where $\gamma= (x-1)/(x+1)$, $K(x)$ and $E(x)$ are the first and
second kinds of complete elliptic integrals, respectively. The
normal current has $dc$ and $ac$ parts which both are zero
in the absence of the spin precession. The spin precession is
sufficient for existence of the $dc$ part and it is independent of
the tilt angle oscillation. Whereas, the $ac$ current is emerged when the
tilt angle $\theta$ oscillates with a small oscillating amplitude
$\vartheta$ and the spin conserving transmission amplitude is
nonzero. Totally, when $\Omega+\omega$ become larger than $2\Delta$ the normal current is
nonzero and different situations will appear depending on the values
of $\Omega$ and $\omega$. For $\Omega<2\Delta$, the $dc$ part of the
current is zero and three discontinuities appear at points
$\Omega+\omega=2\Delta$, $\omega=2\Delta$ and
$\omega-\Omega=2\Delta$. Indeed such discontinuities appear in the
normal current when the gained or lost energy during the interaction
of the single particles with the spin dynamics are almost equal to
the superconducting gap. At these points a resonance occurs in the
junction. Moreover for $\Omega>2\Delta$, there is a $dc$ normal
current in the junction and the discontinuities appear at points
$\Omega-\omega=2\Delta$, $\omega=2\Delta$ and
$\omega-\Omega=2\Delta$.

%%%%%%%%%%%%%%%%%%%%%%%%%%%%%%%%%%%%%%%%%%%%%%%%%%%%%%%%%%%%%%%%%%%%
%%%%%%%%%%%%%%%%%%%%%%%%   Josephson current   %%%%%%%%%%%%%%%%%%%%%

{\it Josephson current} - To calculate the Josephson current we define
a different retarded potential as
\begin{equation}\label{X-definition}
X^{\sigma\sigma'}_{\rho\rho'}(t-t')=-i\Theta(t-t')\left\langle\left[a_{k,k'}^{\sigma\sigma'}(t)
,a_{p,p'}^{\rho\rho'}(t')\right]\right\rangle\ .
\end{equation}
As in the normal case, in the absence of the spin dependent interactions
inside the leads the retarded potential (\ref{X-definition}) could be written as
$X^{\sigma\sigma'}_{\rho\rho'}(t-t')=\sigma\sigma'\delta_{\sigma,-\rho}\delta_{\sigma',-\rho'}X_{ret}(t-t')$,
where $\sigma,\sigma'=\pm 1$. The associated Matsubara potential
reads
\begin{equation}\label{X-Matsobara}
\mathcal{X}(i\omega_n)=\frac{1}{\beta}\sum_{iq}\mathcal{F}_R^{\dagger}(k,iq)\mathcal{F}_L(k',iq-i\omega_n)\
,
\end{equation}
where $\mathcal{F}_R$ and $\mathcal{F}_L$ are the anomalous Green's
functions in the leads \cite{Mahan1990}. The retarded potential is
obtained by analytical continuation. Implementing the real part of
the retarded potential
\begin{equation}\label{Realpart-X}
\Re\left(\sum_{k,k'}X_{ret}(x)\right)=\left\{
\begin{array}{cl}
\pi N_LN_R\Delta K(x)\;& x<1
\\
\pi N_LN_R\frac{\Delta}{x}K(\frac{1}{x})\;& x>1
\end{array}\right.
\end{equation}
and defining ${\cal J}(x)=\Re\left(\sum_{k,k'}X_{ret}(x)\right)/\pi N_LN_R$,
the final form of the Josephson current is given by:
\begin{eqnarray}
&\nonumber i^J(t)=\big[2(T_0^2-T_{||}^2){\cal J}(0)
-2T_{\bot}^2{\cal J}\left(\frac{\Omega}{2\Delta}\right)\\
&-T_{\bot}T_{||}{\bar{\cal J}}\left(\frac{\Omega}{2\Delta},
\frac{\omega}{2\Delta}\right)\vartheta\cos\omega t\big]\sin\chi
\label{Josephson-current}
\end{eqnarray}
where $i^J=I^J/2\pi e N_LN_R$, $\chi=\chi_R-\chi_L$ is the phase difference between
superconducting leads and
\begin{eqnarray}\label{amplitude-josephson-current}
&{\bar{\cal J}}\left(\frac{\Omega}{2\Delta},\frac{\omega}{2\Delta}\right)=2{\cal J}(0)
-2{\cal J}\left(\frac{\Omega}{2\Delta}\right)
+2{\cal J}\left(\frac{\omega}{2\Delta}\right) \nonumber\\&
-{\cal J}\left(\frac{\omega+\Omega}{2\Delta}\right)
-{\cal J}\left(\frac{\omega-\Omega}{2\Delta}\right).
\end{eqnarray}
In the special case of $\vartheta=0$, which corresponds to the
$h_0=0$, the spin has only a precessing motion about the $z$ axis
without tilt angle oscillation. In this case the $dc$ Josephson
current $i^J=2\big[(T_0^2-T_{||}^2){\cal J}(0) -T_{\bot}^2{\cal
J}\left(\frac{\Omega}{2\Delta}\right)]\sin{\chi}$ is generated
through the junction. Indeed, the interaction of the quasiparticles
with the spin precession affects the $dc$ Josephson current and
causes to appear a divergence at $\Omega=2\Delta$, when the quantum
of precession is close to the superconducting gap.

For $\vartheta\neq 0$ and $\theta_0\neq \frac{n\pi}{2}$, depending
on the values of ${\bar{\cal
J}}\left(\frac{\Omega}{2\Delta},\frac{\omega}{2\Delta}\right)$, an
$ac$ Josephson current is generated through the junction. This
modulation of the Josephson current can be used for single spin
detection. At $\Omega=0$, the parameter ${\bar{\cal J}}$ vanishes
for arbitrary values of $\omega$ and the Josephson current
$i^J=2(T_0^2-T_{||}^2){\cal J}(0)\sin\chi$ is time-independent. In
Fig. (\ref{amplitude-josephson}), we have shown the density plot of
the amplitude of $ac$ Josephson current ($\bar{\cal {J}}$) versus
$\frac{\omega}{2\Delta}$ and $\frac{\Omega}{2\Delta}$.

For $\frac{\Omega}{2\Delta}<1$ the amplitude of the $ac$ Josephson current
diverges at points $\frac{\Omega+\omega}{2\Delta}=1$, $\frac{\omega}{2\Delta}=1$
and $\frac{\omega-\Omega}{2\Delta}=1$. Also, as it is clearly observed,
$\bar{\cal {J}}$ changes its sign at two values of $\frac{\omega}{2\Delta}$
around one. However, for $\frac{\Omega}{2\Delta}>1$ two divergencies emerge
at points $\frac{\omega}{2\Delta}=1$ and $\frac{\omega-\Omega}{2\Delta}=1$. The
amplitude vanishes and changes sign at two points around the
$\frac{\omega-\Omega}{2\Delta}=1$. These divergencies appear when the
absorbed or emitted energy by the quasiparticles, due to interaction
with the spin dynamics of SMM are close to the superconducting gap.
At these situations, the resonance condition takes place in the
junction and the current flowing through the junction diverges. So,
tuning the values of the $\Omega$ and $\omega$ close to a resonance
condition may leads to pumping a Josephson current trough the
junction.

{\it Adiabatic limit} - In the adiabatic limit ($\frac{\Omega}{2\Delta}\ll
1$, $\frac{\omega}{2\Delta}\ll 1$), when the evolution in the system are
very slow compared to the dwell time of the quasiparticles, there is
no normal current flowing through the junction and the Josephson
current is simplified as
\begin{eqnarray}
&\nonumber i^J(t)=\pi\Delta[(T_0^2-T_S^2)
-T_{\bot}^2\frac{\Omega}{8\Delta}
+T_{\bot}T_{||}\frac{\Omega}{8\Delta}\vartheta\cos\omega
t]\sin\chi.\\ \label{Adiabatic-Josephson-current}
\end{eqnarray}
Surprisingly, in the adiabatic limit the magnitudes of the {\it dc}
and {\it ac} Josephson currents depend linearly on the precession
frequency $\Omega$ and they are independent of the tilt angle
oscillation frequency $\omega$.

%%%%%%%%%%%%%%%%%%%%%%%%%%%%%%%%%%%%%%%%%%%%%%%%%%%%%%%%%%%%%%%%%%%%
%%%%%%%%%%%%%%%%%%%%%%%%   Conclusion   %%%%%%%%%%%%%%%%%%%%%%%%%%%%

{\it Conclusion} -We introduced a charge current pump, a Josephson
junction consisting of a single molecular magnet with spin nutation
embedded between two spin singlet superconducting leads. The spin
nutation, spin precession combined with an oscillation about it, is
generated by applying a time dependent magnetic field to the single
molecular magnet. The simultaneous effects of precession and
oscillation cause to pump an {\it ac} normal and Josephson currents
through the system. Varying the magnetic field enables us to control
the magnitudes of the pumped currents. At resonance conditions some
discontinuities and divergencies emerge in the normal and Josephson
currents. Such behaviors appear due to the interaction of the
quasiparticles with spin dynamics of the single molecular magnet and
take place when the absorbed/emitted energy during the transferring
is in the range of the superconducting gap.

The long relaxation times of the SMMs and small size of the
junctions consisting of them make them favorable to use in molecular
spintronics and quantum computing. Our introduced system have
interesting properties and would be important from practical point
of view. The modulation of the Josephson current by applied magnetic
filed can make possible the single spin detection. Moreover, tuning
the applied magnetic filed to bring the system close to a resonance
condition enable us to pump $dc$ and $ac$ normal and Josephson
currents trough the system in a controllable way.

%%%%%%%%%%%%%%%%%%%%%%%%%%%%%%%%%%%%%%%%%%%%%%%%%%%%%%%%%%%%%%%%%%%%%%
%%%%%%%%%%%%%%%%%%%%%%%%%%%%%%%%%%%%%%%%%%%%%%%%%%%%%%%%%%%%%%%%%%%%%%

%\begin{acknowledgments}
BA and JA acknowledge Shahrood University of Technology where the initial
parts of this work was done.
%\end{acknowledgments}

\end{document}